\documentclass[12pt,english]{article}
\usepackage{tikz}
\usepackage{mathptmx}
\usepackage{geometry}
\usepackage{xcolor}
\usepackage{array}
\usepackage{bbding}
\usepackage{multirow}
\usepackage[fleqn]{amsmath}
\usepackage{amssymb}
\usepackage{amsthm}
\usepackage{graphicx}
\usepackage{natbib}
\usepackage{lscape}
\usepackage[hyphens]{url}
\usepackage[hidelinks]{hyperref}
\usepackage{comment}
\usepackage{bm}
\usepackage[title]{appendix}
\usepackage{longtable}
\usepackage{mathtools}
\usepackage{chngcntr}
\usepackage{authblk}
\usepackage{babel}
\usepackage{dsfont}
\usepackage{subcaption}
\usepackage{fancyhdr}
\usepackage{cleveref}
\usepackage{abstract}

\makeatletter
\allowdisplaybreaks
\date{}

\makeatletter
\def\blfootnote{\xdef\@thefnmark{}\@footnotetext}
\makeatother

\makeatletter
\def\titlepageext{
	\begin{center}	
	{\parindent0pt
		\rule{0.9\textwidth}{1pt}
		\begin{minipage}[t]{0.25\textwidth}
			\small {\it Keywords:}\\
			\keyword
		\end{minipage}%
		\hspace{3mm}
		\begin{minipage}[t]{0.6\textwidth}
			\small \abstract
		\end{minipage}%
		
		\rule{0.9\textwidth}{2pt}
	}
	\end{center}

	\blfootnote{* Corresponding author. E-mail address: \href{mailto:\corresemail}{\corresemail}.}
}
\makeatother

\usepackage[mathlines, pagewise]{lineno}
\usepackage{etoolbox} 

\newcommand*\linenomathpatchAMS[1]{%
	\expandafter\pretocmd\csname #1\endcsname {\linenomathAMS}{}{}%
	\expandafter\pretocmd\csname #1*\endcsname{\linenomathAMS}{}{}%
	\expandafter\apptocmd\csname end#1\endcsname {\endlinenomath}{}{}%
	\expandafter\apptocmd\csname end#1*\endcsname{\endlinenomath}{}{}%
}

\expandafter\ifx\linenomath\linenomathWithnumbers
\let\linenomathAMS\linenomathWithnumbers
\patchcmd\linenomathAMS{\advance\postdisplaypenalty\linenopenalty}{}{}{}
\else
\let\linenomathAMS\linenomathNonumbers
\fi

\linenomathpatchAMS{gather}
\linenomathpatchAMS{multline}
\linenomathpatchAMS{align}
\linenomathpatchAMS{alignat}
\linenomathpatchAMS{flalign}

\nolinenumbers

\title{Noise-induced stop-and-go traffic dynamics:\\ Modelling and control}

\author[a]{Raphael Korbmacher}
\author[b]{Parthib Khound}
\author[a,*]{Antoine Tordeux}
\author[c]{Frank Gronwald}
\affil[a]{University of Wuppertal, Gaußstraße 20, 42119 Wuppertal, Germany}
\affil[b]{Indian Institute of Technology Bombay, 400076 Mumbai, India}
\affil[c]{University of Siegen, Hölderlinstraße 3, 57076 Siegen, Germany}

\def\corresemail{tordeux@uni-wuppertal.de}

\def\abstract{Stop-and-go waves in traffic flow are captivating collective phenomena with important safety and environmental consequences. While classical theories attribute these oscillations to linear instabilities caused by reaction delays and inertia, this study explores an alternative stochastic perspective. Using a linearly stable car-following model, we show that white Gaussian noise in the measurement of the inter-vehicle distance can destabilise the flow, inducing a phase transition to periodic stop-and-go dynamics via a nonlinear instability mechanism. Furthermore, we demonstrate that a simple linear transformation of the model, which amplifies the system response while introducing a positive acceleration bias, can counteract noise-induced effects and restores the stability of uniform traffic flow. These findings, supported by numerical simulations, aim to provide new insights into the modelling and control of oscillatory traffic dynamics.}

\def\keyword{Road traffic flow\\ Stop-and-go wave\\ Noise-induced instability\\ Nonlinear instability\\ Stabilisation}

\begin{document}
\maketitle
\titlepageext

\section{Introduction}

\begin{figure}[!ht]
    \centering\vspace{-10mm}\footnotesize
    \includegraphics[width=.8\textwidth]{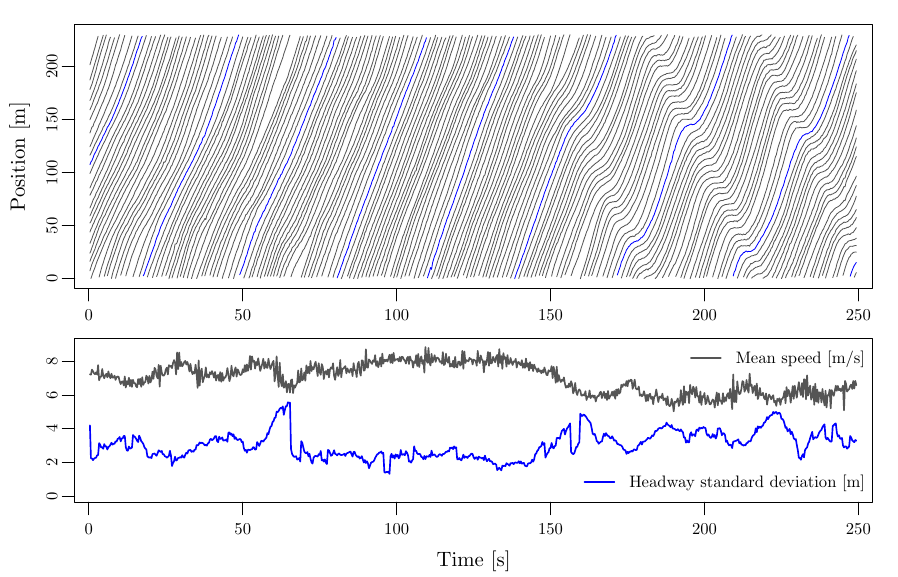}
    \caption{Last 250\ s of the trajectories of 22 vehicles on a single-lane circular track, starting from a uniform configuration \cite{sugiyama2008traffic}. After some time, a stop-and-go wave spontaneously emerges, leading to a decrease in the average speed and an increase in the standard deviation of the inter-vehicle gap.}
    \label{fig:TrajExp}  
\end{figure}

Most car drivers have experienced stop-and-go waves on highways, where traffic jams emerge seemingly for no apparent reason, forcing drivers to repeatedly slow down and speed up.
These waves are striking collective phenomena observed worldwide. 
Beyond their scientific intrigue, stop-and-go dynamics pose significant safety and environmental concerns, as the repeated acceleration and deceleration phases increase fuel consumption and pollutant emissions \cite{li2014stop}.
The term \emph{stop-and-go waves} dates back to Duckstein in the late 1960s \cite{duckstein1967control}. 
More recent expressions, such as \emph{phantom jam} \cite{sugiyama2008traffic} or \emph{jamiton} \cite{flynn2009self}, emphasize their spontaneous appearance without any obvious external trigger.
Stop-and-go waves have been frequently documented in empirical studies. 
One of the first controlled scientific experiments was conducted in the late 2000s using 22 vehicles on a single-lane circular track of 230\ m \cite{sugiyama2008traffic}, as shown in Fig.~\ref{fig:TrajExp}.
Starting from a homogeneous configuration, stop-and-go waves emerge after a while, leading to self-sustained oscillatory dynamics.

Stop-and-go waves are classically understood in the literature as linear instabilities in deterministic modelling frameworks \cite{wilson_CarfollowingModelsFifty_2011}. 
However, despite over seven decades of research, the emergence of self-sustaining waves in traffic flow remains to be not completely understood. 
Many recent experiments reveal that even advanced driver assistance systems, such as adaptive cruise control (ACC), exhibit major stability issues \cite{milanes2014modeling,gunter2020commercially,ciuffo2021requiem,li2021car,makridis2021openacc}. 
Understanding and mitigating traffic oscillations remains a complex challenge, with ongoing efforts exploring novel approaches \citep{korbmacher2025understanding}. For instance, recent studies have leveraged reinforcement learning techniques to dissipate traffic waves and stabilise traffic flow \citep{kreidieh2018dissipating,lee2025traffic, jang2024reinforcement,korbmacher2025emergent}. 
In fact, many factors can perturb the stability and cause the flow to collectively oscillate. 
Stability requires the agents to be accurate and reactive. 
However, driver behaviour, vehicle control, environmental perception, and dynamic aspects are intrinsically subject to fluctuations, inaccuracies, reaction time, and latency. 

In this contribution, we first review, in Section~\ref{Sec:Review}, the state of the art in traffic instabilities for both deterministic and stochastic car-following models. 
We then present, in Section~\ref{Sec:Simulation}, an original stochastic approach showing that white Gaussian noise in the measurement of the distance to the predecessor within a stable nonlinear car-following model can destabilize the system, inducing a phase transition from laminar to periodic dynamics. 
The uniform solution remains stable for small perturbations, while an oscillatory solution with stop-and-go waves becomes spontaneously stable when perturbations exceed a critical threshold. 
Thus, nonlinear instabilities driven solely by noise can trigger oscillatory dynamics in an otherwise unconditionally linearly stable model, provided the noise amplitude is sufficiently large. 
Some simulations reproducing Sugiyama’s experiment \cite{sugiyama2008traffic} illustrate these effects. 
Furthermore, in Section~\ref{Sec:Stab}, we show that a simple linear transformation of the model, amplifying its response and adding a positive acceleration bias, can mitigate the noise effects and substantially enhance the stability of the uniform flow. 
We interpret this noise-induced collective behaviour using an analogy to the Kapitza inverted pendulum and the concepts of stochastic resonance and stochastic stabilisation in Section~\ref{Sec:Dis}.
Finally, Section~\ref{Sec:Ccl} provides a summary and concluding remarks.

\section{Reviewing car-following models and traffic instabilities\label{Sec:Review}}

The first studies pointing out instability phenomena in single-file traffic systems date back to the early 1950s, with the pioneering work of Reuschel \cite{reuschel1950fahrzeugbewegungen} and, a few years later, Pipes et al.\ \cite{pipes_OperationalAnalysisTraffic_1953}. 
These groundbreaking studies examined the behaviour of road vehicles using first- and second-order models without delay. 
By the late 1950s, research expanded to include delays in linear models, as in \cite{kometani1958stability,herman_TrafficDynamicsAnalysis_1959,chandler1958traffic,kishi1960traffic}, or in nonlinear models, as in \cite{newell1961nonlinear,gazis1961nonlinear}.
These prolific years laid the foundations for many fundamental stability concepts using linear algebra \cite{reuschel1950fahrzeugbewegungen}, Laplace transforms \cite{pipes_OperationalAnalysisTraffic_1953,kometani1958stability,herman_TrafficDynamicsAnalysis_1959}, alongside with Fourier analysis \cite{chandler1958traffic}, see \cite{orosz2010traffic} for a survey.
Conceptually, researchers differentiate between platoon stability, relevant for finite systems, and string stability, which applies to circular tracks or infinite systems and is generally more restrictive. 
The notions of damped stability, characterised by oscillatory behaviour, and overdamped stability, with lack of oscillations, were also formalised during this period \cite{kishi1960traffic}. 
An important physical insight is the detrimental effect of delay and inertia on flow stability. 
Unstable models amplify perturbations, leading to the emergence of traffic waves. 
This represented a significant improvement over earlier first-order fluid-dynamics macroscopic models \cite{lighthill1955kinematic,richards1956shock}, which are inherently stable and fail to capture such instability phenomena.

Another fundamental behavioural concept emerged in the late 1950s: maintaining a constant time gap with the preceding vehicle. 
The existence of a phenomenological relationship between speed and inter-vehicle distance can in fact be traced back to some of the earliest traffic measurements in the 1930s \cite{greenshield1934}. This relationship was later referred to in the literature as the \emph{California Code} \cite{chandler1958traffic} and physically interpreted in terms of safe braking distances \cite{gipps_BehaviouralCarfollowingModel_1981}.
Today, the constant time-gap strategy, or more generally, the assumption of an equilibrium relationship between speed and spacing, is embedded in most modern car-following models, either through optimal velocity (OV) functions \cite{bando_DynamicalModelTraffic_1995,jiang_FullVelocityDifference_2001} or through desired time-gap parameters \cite{treiber_CongestedTrafficStates_2000,tordeux2010adaptive}. Moreover, standards for adaptive cruise control systems recommend maintaining a constant time gap to the predecessor, typically ranging from 0.8 to 2.2~s in ISO standards \cite{iso15622}.

\subsection{Car-following models and their stability conditions}

After its initial prominence in the 1950s and early 1960s, stability analysis experienced a period of decline, interrupted only in the 1970s by studies on multi-anticipatory models \cite{bexelius1968extended}. 
Interest resurged in the late 1990s with the introduction of the Optimal Velocity Model (OVM) \cite{bando_DynamicalModelTraffic_1995}. Numerical simulations revealed that, for certain parameter regimes, nonlinear optimal velocity functions can exhibit phase transitions between two stable stationary states: a homogeneous traffic state and a heterogeneous state characterized by backward-propagating stop-and-go waves. 

\paragraph{Optimal velocity models}
Denoting as $x_n(t)$ the position of the vehicle $n$ at time $t$, $v_n(t)=\text d x_n(t)/\text dt$ its velocity, and $\Delta x_n(t)=x_{n+1}(t)-x_n(t)$ the distance to the predecessor,
the OV model is given by the differential equation \cite{bando_DynamicalModelTraffic_1995}:  
\begin{equation}
\frac{\text dv_n(t)}{\text dt} = \frac{1}{\tau}\big[V(\Delta x_n(t)) - v_n(t)\big],\qquad\tau>0,\quad V\in C^1(\mathbb R,\mathbb R_+).
\label{eq:OVM}
\end{equation}
In the \emph{Full Velocity Difference} (FVD) model, the dynamics is
\begin{equation}
\frac{\text dv_n(t)}{\text dt} = \frac{1}{\tau_1}\big[V(\Delta x_n(t)) - v_n(t)\big] +\frac{1}{\tau_2}\Delta v_n(t),\qquad\tau_1,\tau_2>0,\quad V\in C^1(\mathbb R,\mathbb R_+),
\label{eq:FVDM}
\end{equation}
where $\Delta v_n(t)=v_{n+1}(t)-v_n(t)$ is the speed difference with the preceding vehicle.
In these models, $V\in C^1(\mathbb R,\mathbb R_+)$ is the optimal velocity function while $\tau,\tau_1,\tau_2>0$ are relaxation time parameters. 
A typical example of OV function is the hyperbolic tangent
\begin{equation}
V(x) = \frac{V_0}2\bigg[1 + \tanh\bigg(\frac{x-\ell}T-A\bigg)\bigg].
\end{equation}
Other shapes include the sigmoid function \cite{newell1961nonlinear}
\begin{equation}
    V(x) = V_0\bigg[1 -
    \exp\bigg(-\frac{x-\ell}T\bigg)\bigg],
\end{equation}
which is also used to model pedestrian dynamics \cite{weidmann1993transporttechnik}, or the part-linear form \cite{newell2002simplified}
\begin{equation}
    V(x) = \text{smin}\bigg\{V_0,\text{smax}\bigg\{0,\frac{x-\ell}T\bigg\}\bigg\},
\end{equation}
where $\text{smin}$ and $\text{smax}$ are smooth minimum and maximum functions.
These OV functions are based on three parameters: $\ell\ge0$ corresponds to the length of the vehicle, $T>0$ is the desired time gap, and $V_0>0$ is the desired (maximal) speed.

\paragraph{Adaptive time gap model}
\def\tmin{T_{\text{min}}}
\def\tmax{T_{\text{max}}}
In contrast to classical OV models that relax the speed to a desired (or optimal) speed function that depends on the gap \cite{bando_DynamicalModelTraffic_1995,jiang_FullVelocityDifference_2001}, the adaptive time gap (ATG) car-following model~\cite{tordeux2010adaptive} is obtained by relaxing the time gap 
\begin{equation}
    T_n(t)=\frac{g_n(t)}{v_n(t)},\qquad g_n(t)=\Delta x_n(t)-\ell,
\end{equation} 
which is the time taken by a vehicle to reach the vehicle in front, assuming that the speed is constant and the speed of the vehicle in front is zero.
The dynamics for the time gap is given by 
\begin{equation*}
\frac{\text dT_n(t)}{\text dt}=\frac1\tau\bigl(T-T_n(t)\bigr),\qquad \tau,T>0, 
\end{equation*} 
where $\tau>0$ 
is a sensitivity parameter (relaxation time) and $ T>0$ is the desired time gap. 
Using the speed and distance variables, the ATG model in its Newtonian form is given by 
\begin{equation}
    \frac{\text dv_n(t)}{\text dt} =\frac{1}{T_n(t)}\biggl[\frac1\tau\left[\Delta x_n(t)-V^{-1}\big(v_n(t)\big)\right]+\Delta v_n(t)\biggr],
    \label{eq:ATG}
\end{equation}
where $V^{-1}(v)=Tv+\ell$ is the inverse of the optimal velocity function which is assumed to be affine. 
In contrast to the FVD model (see Eq.~\eqref{eq:FVDM}), the ATG is strongly nonlinear because one of the relaxation times is not a constant parameter, but rather the time gap variable.
As we will see in the following, the ATG model is unconditionally linearly stable for any $\tau,T>0$~\cite{khound2023extending}.
However, the introduction of white noise in the dynamics, and more precisely in the measurement of the gap to the predecessor, induces a phase transition to collective oscillatory behaviour with stop-and-go waves.

\paragraph{Intelligent driver model}

The \emph{intelligent driver} (ID) model is a well-known nonlinear car-following model introduced in the early 2000's \cite{treiber_CongestedTrafficStates_2000}, given by
\begin{equation}
    \text dv_n(t) = \alpha \bigg( 1 - \bigg(\frac{f(v_n(t),v_{n+1}(t))}{\Delta x_n(t)-\ell}\bigg) ^ 2 - \Big(\frac{v_n(t)}{V_0}\Big) ^ 4\bigg)\text dt,
\label{eq:IDM}
\end{equation}
where
\begin{equation}
    f(v_n,v_{n+1}) = T v_n - v_n \frac{v_{n+1} - v_n}{2\sqrt{\alpha\beta}},\qquad \alpha,\beta>0.
\label{eq:SIDM2}
\end{equation}
Here, $\alpha$ denotes the maximum acceleration and $\beta$ the comfortable deceleration.

\paragraph{Linear stability conditions}

Consider the general car-following model   
\begin{equation}
    \frac{\text dv_n(t)}{\text dt}= F\big(\Delta x_n(t),v_n(t),v_{n+1}(t))
\end{equation}
where $F\in C^1(\mathbb R^3,\mathbb R)$. 
Suppose that there exists an equilibrium solution $(v^\ast,\Delta x^\ast)$ such that \begin{equation}
    F(\Delta x^\ast,v^\ast,v^\ast)=0.
    \end{equation}
Linearising the speed dynamics around the uniform equilibrium solution gives
\begin{equation}
    \frac{\text d\tilde v_n(t)}{\text dt} = a\Delta\tilde x_n(t) + b \tilde v_n(t) + c \tilde v_{n+1}(t)
    \label{eq:ModelLinear}
\end{equation}
where
\begin{equation}
    \Delta\tilde x_n(t)= \Delta x_n(t)-\Delta x^\ast,\qquad \tilde v_n(t) = v_n(t)-v^\ast,
\end{equation}
while, with $F:(x,y,z)\mapsto F(x,y,z)$,
\begin{equation}
    a=\frac{\partial F}{\partial x}(\Delta x^\ast,v^\ast,v^\ast),\qquad b=\frac{\partial F}{\partial y}(\Delta x^\ast,v^\ast,v^\ast),\qquad c=\frac{\partial F}{\partial z}(\Delta x^\ast,v^\ast,v^\ast).
\end{equation}
It is easy to check that the characteristic equation of the linear dynamics \eqref{eq:ModelLinear} reads
\begin{equation}
    \zeta_\theta^2+\zeta_\theta(b+c\exp(\mathrm i \theta))+a(1-\exp(\mathrm i \theta))=0,\qquad \zeta_\theta\in\mathbb C, \quad\theta\in(0,2\pi).
    \label{eq:EquaCharac}
\end{equation}
The characteristic equation is a second-order polynomial with complex coefficients.
The system is linearly stable if the real part of the roots $\zeta_\theta$ are nonpositive for all $\theta\in(0,2\pi)$. 
This holds true if \cite{tordeux2012linear}, \cite[Chap.~15]{treiber2013traffic}
\begin{equation}
    a>0,\qquad b<0,\qquad\text{and}\qquad b^2-c^2 \ge 2a.
    \label{eq:stable}
\end{equation}
These conditions can be recovered by using general stability conditions for complex polynomials in \cite[Th.~3.2]{frank1946zeros} which are generalisation of the Hurwitz conditions. 
In addition, the system with $c>0$ is overdamped stable, i.e., it does not present oscillation during the stabilisation phase, if \cite{khound2020local}
\begin{equation}
    a>0,\qquad b<0,\qquad\text{and}\qquad \frac ac+b+c\le0.
    \label{eq:ovdstable}
\end{equation}
The conditions are sufficient for any finite system and become exact for infinite systems.
Note that the overdamped stability condition is stronger than the simple stability, i.e., \eqref{eq:ovdstable}~$\Rightarrow$~\eqref{eq:stable}. 
Indeed both conditions \eqref{eq:stable} and \eqref{eq:ovdstable}  imply that $b+c\le0$. In addition, the last condition of \eqref{eq:stable} is 
\begin{equation}
    b+c\ge \frac{2a}{b-c}
\end{equation}
while the last condition of \eqref{eq:ovdstable} is 
\begin{equation}
b+c\le -\frac ac.
\end{equation}
Therefore \eqref{eq:ovdstable}~$\Rightarrow$~\eqref{eq:stable} if and only if $-a/c\ge 2a/(b-c)$ which is $b+c\le0$ after simplifications.

For example, for the FVD model~\eqref{eq:FVDM} with $V'(\Delta x^\ast)=1/T$, the partial derivatives at equilibrium are given by
\begin{equation}
    a_\text{FVD}=\frac{1}{T\tau_1},\qquad b_\text{FVD}=-\frac1\tau_1-\frac1{\tau_2},\qquad c_\text{FVD}=\frac1{\tau_2}.
\end{equation}
The linear stability condition \eqref{eq:stable} is \cite[Chap.~15]{treiber2013traffic}
\begin{equation}
    \tau_1,\tau_2,T>0,\qquad \frac{2\tau_1\tau_2}{2\tau_1+\tau_2} \le T,
    \label{eq:CSFVD}
\end{equation}
while the overdamped stability condition \eqref{eq:ovdstable} is simply given by 
\begin{equation}
    \tau_1>0,\qquad0<\tau_2\le T.
    \label{eq:CSFVD2}
\end{equation}
Note that both stability conditions \eqref{eq:CSFVD} and \eqref{eq:CSFVD2} hold true for $\tau_2=T>0$.

The partial derivatives at equilibrium for ID model \eqref{eq:IDM} at the limit $V_0\to\infty$ are
\begin{equation}
    a_\text{ID}=\frac{2\alpha}{Tv^\ast},\qquad
    b_\text{ID}=-\frac{2\alpha}{v^\ast}-\frac1{T}\sqrt{\frac \alpha\beta},\qquad
    c_\text{ID}=\frac1{T}\sqrt{\frac \alpha\beta}.
\end{equation}
With these, the string stability condition \eqref{eq:stable} is satisfied when
\begin{equation}
    \label{IDB_stringStability}
    \alpha,\beta,v^\ast,T>0,\qquad \frac{v^\ast}{\alpha} \bigg(1-\sqrt{\frac{\alpha}{\beta}}\, \bigg)\le T\,.
\end{equation}
Further, the overdamped stability criterion \eqref{eq:ovdstable} is fulfilled if and only if
\begin{equation}
    \label{IDB_OverdampedStability}
    v^\ast,T>0,\qquad 0<\beta\le \alpha 
\end{equation}
The above condition will make the left side of \eqref{IDB_stringStability} non-positive, so \eqref{IDB_OverdampedStability}$\Rightarrow$\eqref{IDB_stringStability}. 

For the ATG model~\eqref{eq:ATG}, the partial derivatives at equilibrium are 
\begin{equation}
\label{ATG_linearized_PD}
    a_\text{ATG}=\frac{1}{T\tau},\qquad 
    b_\text{ATG}=-\frac1\tau-\frac1T,\qquad 
    c_\text{ATG}=\frac1T,
\end{equation}
and the stability conditions \eqref{eq:stable} and \eqref{eq:ovdstable} hold true for any $\tau,T>0$.
The ATG model is unconditionally linearly string and overdamped stable, hence its prevalence for automated driving systems \cite{khound2023extending}.

Linear stability analysis underscores the central role of inertia and nonlinear mechanisms in explaining stop-and-go traffic dynamics. 
Hopf bifurcation allows to characterise the phase transition from uniform (laminar) dynamics to oscillatory behaviour with stop-and-go waves \cite{orosz2004global, orosz2010traffic, tordeux2012linear}. 
Different perturbation dynamics can be identified, including convective upstream, stationary, and convective downstream modes~\cite{wilson_CarfollowingModelsFifty_2011}.
Subsequent research with reductive perturbation methods derives macroscopic wave patterns given by the Korteweg-de Vries (KdV) equation \cite{komatsu1995kink} and the modified KdV equation \cite{nagatani1998modified} from optimal velocity models. 
Additionally, nonlinear models can demonstrate metastable behaviour, exhibiting multiple stationary dynamics depending on initial conditions \cite{tomer2000presence}.
In the literature, these results paved the way for the analysis of stop-and-go wave phenomena and phase transitions in traffic flow based on a linear instability of inertial, i.e., second-order, nonlinear models \cite{gasser2004bifurcation,orosz2010traffic}.

\subsection{Stochastic models}

The inertia-induced linear instability of classical car-following models explains well the phenomena of spontaneous perturbation amplification recently observed in experiments with ACC-equipped vehicles \cite{gunter2020commercially,makridis2021openacc}. 
Indeed, the response times of the ACC-systems seem too large to fulfil basic linear stability conditions \cite{makridis2019response}. 
Another alternative modelling approach of traffic instabilities relies on stochastic aspects. 
Pioneering work in this area have been done using discrete cellular automata and interacting particle systems \cite{BarlovicSSS98,Ke-Ping_2004,kaupuvzs2005zero,huang2018instability}, \cite[Chap. 8.1]{Schadschneider:1564539}. 
More recently, continuous modelling approaches based on stochastic differential equation systems incorporate additive noise \cite{treiber2009hamilton,wagner2011time,laval2014parsimonious}. 
A general stochastic car-following model is given by
\begin{equation}
\left\{\begin{aligned}
    \text dx_n(t) &= v_n(t)\,\text dt\\
    \text dv_n(t) &= F\big(\Delta x_n(t),v_n(t),\Delta v_{n+1}(t)\big)\text dt +\sigma \text d W_n(t),
    \end{aligned}\right.
\end{equation}
where $F\in C^1(\mathbb R^3,\mathbb R)$ is a continuously differentiable nonlinear function governing the motion, while $\sigma \text d W_n $ is a white noise with $W_n$ a standard Wiener process and $\sigma\ge0$ the noise amplitude.

\paragraph{Linear or near-linear stochastic models}
Linear or near-linear stochastic models derived from optimal velocity car-following models are given by \cite{treiber2009hamilton,wagner2011time,laval2014parsimonious,wang2020stability,friesen2021spontaneous}:
\begin{eqnarray}
    \text dv_n(t)=-\kappa\big[V\big(\Delta x_n(t)\big) - v_n(t)\big]\text dt+c \Delta v_n(t)\text dt+\sigma \text d W_n(t),\qquad \kappa<0,\quad c\ge0,
    \label{SOVM}
\end{eqnarray}
where $V$ is the optimal velocity function. 
We recover a stochastic version of the OV model for $c=0$, whereas a stochastic FVD model is recovered for $c>0$.
Systems with linear OV functions $V$ are ergodic multidimensional Ornstein-Uhlenbeck processes having a stable dynamics with a single invariant Gaussian distribution \cite{wang2020stability,friesen2021spontaneous}. 
More realistic models have nonlinear OV functions $V$ (see, e.g., \cite{treiber2009hamilton}).
Another example of near-linear stochastic models is the inertial car-following model \cite{tomer2000presence}
\begin{equation}
    \text dv_n(t) = K \bigg(1 - \frac{2v_n(t) T + \ell}{\Delta x_n(t)-\ell}\bigg)\text dt + \frac{Z^2(\Delta v_n(t))}{ 2\Delta x_n(t)}\text dt
    -2Z(v_n(t) - V_0)\text dt + \sigma d W_n(t),\quad K>0,
    \label{Tomer}
\end{equation}
where $Z(x)=(x+|x|)/2$ is the positive part of $x$ and $T>0$ the desired time gap. 
These models include nonlinear components through the OV function or sensitivity coefficients. 
However, they behave mainly as linear models when stable and do not describe a phase transition as the noise volatility increases, except in subcritical regimes \cite{treiber2017intelligent}. 
In fact, the stability condition is not affected by white noise with linear models. 

\paragraph{Nonlinear stochastic models}
The stochastic extension of the ID model (SID), studied more recently \cite{treiber2017intelligent}, is given by
\begin{equation}
    \text dv_n(t) = \alpha \bigg( 1 - \bigg(\frac{f(v_n(t),v_{n+1}(t))}{\Delta x_n(t)-\ell}\bigg) ^ 2 - \Big(\frac{v_n(t)}{V_0}\Big) ^ 4\bigg)\text dt + \sigma \text dW_n(t).
\label{eq:SIDM}
\end{equation}
In contrast to the previous linear and near-linear models, the SID model is strongly nonlinear. It describes a phase transition resulting from a subcritical instability as the noise volatility increases \cite{treiber2017intelligent}. 
This transition is caused by both the nonlinearity and the underlying linear instability of the deterministic intelligent driver model.
It only occurs in the vicinity of the critical parameter setting.
Note that state-dependent noise models based on the the Cox-Ingersoll-Ross process also exhibit phase transition as the noise volatility increases, see, for example, \cite{ngoduy2019langevin,xu2019analysis}. 
In these models, the transition also occurs from a linear instability because the noise function depends on the system state, rather than the form of the OV function as it does in deterministic car-following models.

\section{Noise-induced phase transition\label{Sec:Simulation}}

In this section, we point out through simulations that introducing white Gaussian noise into gap measurements of the ATG model leads to nonlinear instability of the uniform solution and the emergence of stop-and-go oscillatory dynamics. 
The ATG model, although unconditionally linearly stable, describes a phenomenon of nonlinear instability when strongly perturbed. 

\subsection{Stochastic ATG model}

In \cite{dufour2025noiseinducedtransitionstopandgowaves}, a noise term is added to the acceleration function of the ATG model.
In this article, we introduce noise into the measurement of the gap. 
Human perception of distances is indeed imperfect. 
Distance sensors also have a limited accuracy, particularly in challenging conditions, e.g., due to bad weather or road curvature. 
We expect the noise to be correlated over time, using, e.g., an Ornstein-Uhlenbeck process, and dependent on the measured distance; estimation is typically more accurate for small distances.
However, to keep the modelling approach minimalistic, we use a simple white noise, independent of the system state and uncorrelated in time. 
Additionally, rather than estimating the noise amplitude empirically, we carry out simulations by varying the noise amplitude range.

\paragraph{Time-continuous stochastic model}
The stochastic ATG model including white noise in the gap is given by 
\begin{equation}
    \mathrm{d}v_n(t) =\frac{\frac1\tau \big[g_n(t)+\sigma\xi_n(t)- Tv_n(t)\big]+\Delta v_n(t)}{T_\varepsilon\big(g_n(t)+\sigma\xi_n(t),v_n(t)\big)}\mathrm{d}t,
    \label{eq:SATG}
\end{equation}
where $\xi_n$ denotes independent standard Gaussian noise uncorrelated in time, i.e., $\langle \xi_n(t)\, \xi_n(t') \rangle = \delta(t-t')$, where $\delta(\cdot)$ denotes the Dirac delta distribution. $\sigma>0$ is the noise amplitude.
Here, the time gap is approximated using the smooth minimum (resp.\ maximum) $T_\varepsilon$ bounded between $\tmin$ and $\tmax$, $0<\tmin<\tmax$,
\begin{equation}
        T_\varepsilon(g,v)= f_\varepsilon\left(\tmin,f_{-\varepsilon}\left(\tmax,\frac{g}{f_{\varepsilon}(0,v)}\right)\right),
\end{equation}
where $f_\varepsilon$ is the \emph{LogSumExp} function given by
\begin{equation}\label{eq:LogSumExp}
   f_\varepsilon(x,y)=\varepsilon \log\big(\exp(x/\varepsilon)+\exp(y/\varepsilon)\big).
\end{equation}
The function $f_\varepsilon(x,y)$ converges to the maximum of $x$ and $y$ as $\varepsilon\to0^+$ and to the minimum as $\varepsilon\to0^-$. In the following, we use $\varepsilon=0.01$.
This smoothing of the dynamics avoids singularities when the vehicles collide due to noise or when the speed is zero. 

\paragraph{Time-discrete stochastic model}
The numerical simulations of the stochastic ATG model are performed using an implicit Euler scheme for the vehicle positions, and an explicit scheme for the vehicle speeds. 
This coupled  numerical scheme is given by
\begin{equation}
\begin{cases}
    ~\displaystyle x_n(t+\delta t)=x_n(t)+\delta t v_n(t+\delta t)\\[2mm]
    ~\displaystyle v_n(t+\delta t) =v_n(t)+\delta t \frac{\frac1\tau \big[g_n(t)+\sigma\xi_n(t)- Tv_n(t)\big]+\Delta v_n(t)}{T_\varepsilon\big(g_n(t)+\sigma\xi_n(t),v_n(t)\big)}. 
\end{cases}
\label{eq:SATGdiscrete}
\end{equation}
The time step of the simulations is defined as $\delta t=0.01$~s. 
In addition, the parameter values are the following: 
$\tau=5$~s, $ T=0.9$~s, $\ell=3$~m, $\tmin=0.1$~s and $\tmax=2.5$~s. 
Here, $\tau=5$~s is chosen based on statistical estimations \cite{tordeux2010adaptive}. 
The parameters $\ell=3$~m and $T=0.9$~s are calibrated to obtain a mean speed approximately equal to 30~km/h, as requested during the experiment by Sugiyama et al.\ \cite{sugiyama2008traffic}, and a propagation speed of the stop-and-go wave backward of approximately 20~km/h, as empirically  observed. In addition, the setting $\tmax=2.5$~s corresponding to the time gap on wave output is used to match the duration of the stop phase. 
The parameter $\tmin$ does not seem to influence the dynamics as long as it remains small and greater than $\delta t$.

\subsection{Simulation results}

\begin{figure}[!ht]
    \centering\vspace{2mm}
    \footnotesize
\begin{tikzpicture}[x=1.5pt,y=1.5pt]
\definecolor{fillColor}{RGB}{255,255,255}
\path[use as bounding box,fill=fillColor,fill opacity=0.00] (0,0) rectangle (247.08, 77.21);
\begin{scope}
\path[clip] ( 5.50, 5.50) rectangle (235.08, 65.21);
\definecolor{drawColor}{RGB}{0,0,0}

\path[draw=drawColor,line width= 0.4pt,line join=round,line cap=round] (123.54, 58.14) --
	(126.50, 58.14) --
	(129.46, 58.11) --
	(132.42, 58.07) --
	(135.36, 58.01) --
	(138.30, 57.93) --
	(141.22, 57.84) --
	(144.12, 57.73) --
	(147.00, 57.60) --
	(149.86, 57.46) --
	(152.69, 57.30) --
	(155.50, 57.13) --
	(158.27, 56.94) --
	(161.00, 56.73) --
	(163.70, 56.51) --
	(166.36, 56.27) --
	(168.98, 56.02) --
	(171.55, 55.75) --
	(174.07, 55.47) --
	(176.54, 55.18) --
	(178.96, 54.87) --
	(181.33, 54.54) --
	(183.64, 54.21) --
	(185.88, 53.86) --
	(188.07, 53.49) --
	(190.19, 53.12) --
	(192.24, 52.73) --
	(194.23, 52.34) --
	(196.14, 51.93) --
	(197.99, 51.51) --
	(199.76, 51.08) --
	(201.45, 50.63) --
	(203.06, 50.19) --
	(204.60, 49.73) --
	(206.05, 49.26) --
	(207.43, 48.78) --
	(208.72, 48.30) --
	(209.92, 47.81) --
	(211.04, 47.31) --
	(212.07, 46.81) --
	(213.01, 46.30) --
	(213.87, 45.79) --
	(214.63, 45.27) --
	(215.30, 44.75) --
	(215.88, 44.22) --
	(216.37, 43.69) --
	(216.77, 43.16) --
	(217.07, 42.63) --
	(217.28, 42.09) --
	(217.40, 41.56) --
	(217.42, 41.02) --
	(217.35, 40.48) --
	(217.19, 39.95) --
	(216.93, 39.41) --
	(216.58, 38.88) --
	(216.14, 38.35) --
	(215.60, 37.82) --
	(214.98, 37.30) --
	(214.26, 36.78) --
	(213.45, 36.26) --
	(212.55, 35.75) --
	(211.57, 35.25) --
	(210.49, 34.75) --
	(209.33, 34.25) --
	(208.08, 33.77) --
	(206.75, 33.29) --
	(205.34, 32.81) --
	(203.84, 32.35) --
	(202.27, 31.90) --
	(200.61, 31.45) --
	(198.88, 31.02) --
	(197.07, 30.59) --
	(195.20, 30.18) --
	(193.24, 29.77) --
	(191.22, 29.38) --
	(189.14, 29.00) --
	(186.98, 28.63) --
	(184.77, 28.27) --
	(182.49, 27.93) --
	(180.15, 27.60) --
	(177.76, 27.28) --
	(175.31, 26.98) --
	(172.82, 26.69) --
	(170.27, 26.42) --
	(167.67, 26.16) --
	(165.04, 25.91) --
	(162.36, 25.68) --
	(159.64, 25.47) --
	(156.88, 25.27) --
	(154.10, 25.09) --
	(151.28, 24.92) --
	(148.43, 24.77) --
	(145.56, 24.64) --
	(142.67, 24.52) --
	(139.76, 24.42) --
	(136.83, 24.34) --
	(133.89, 24.27) --
	(130.94, 24.22) --
	(127.98, 24.18) --
	(125.02, 24.17) --
	(122.06, 24.17) --
	(119.09, 24.18) --
	(116.14, 24.22) --
	(113.18, 24.27) --
	(110.24, 24.34) --
	(107.32, 24.42) --
	(104.40, 24.52) --
	(101.51, 24.64) --
	( 98.64, 24.77) --
	( 95.80, 24.92) --
	( 92.98, 25.09) --
	( 90.19, 25.27) --
	( 87.44, 25.47) --
	( 84.72, 25.68) --
	( 82.04, 25.91) --
	( 79.40, 26.16) --
	( 76.81, 26.42) --
	( 74.26, 26.69) --
	( 71.76, 26.98) --
	( 69.32, 27.28) --
	( 66.92, 27.60) --
	( 64.59, 27.93) --
	( 62.31, 28.27) --
	( 60.09, 28.63) --
	( 57.94, 29.00) --
	( 55.85, 29.38) --
	( 53.83, 29.77) --
	( 51.88, 30.18) --
	( 50.00, 30.59) --
	( 48.20, 31.02) --
	( 46.47, 31.45) --
	( 44.81, 31.90) --
	( 43.24, 32.35) --
	( 41.74, 32.81) --
	( 40.33, 33.29) --
	( 38.99, 33.77) --
	( 37.75, 34.25) --
	( 36.59, 34.75) --
	( 35.51, 35.25) --
	( 34.52, 35.75) --
	( 33.63, 36.26) --
	( 32.82, 36.78) --
	( 32.10, 37.30) --
	( 31.47, 37.82) --
	( 30.94, 38.35) --
	( 30.49, 38.88) --
	( 30.14, 39.41) --
	( 29.89, 39.95) --
	( 29.72, 40.48) --
	( 29.65, 41.02) --
	( 29.68, 41.56) --
	( 29.79, 42.09) --
	( 30.00, 42.63) --
	( 30.31, 43.16) --
	( 30.70, 43.69) --
	( 31.19, 44.22) --
	( 31.78, 44.75) --
	( 32.45, 45.27) --
	( 33.21, 45.79) --
	( 34.06, 46.30) --
	( 35.01, 46.81) --
	( 36.04, 47.31) --
	( 37.16, 47.81) --
	( 38.36, 48.30) --
	( 39.65, 48.78) --
	( 41.02, 49.26) --
	( 42.48, 49.73) --
	( 44.01, 50.19) --
	( 45.63, 50.63) --
	( 47.32, 51.08) --
	( 49.09, 51.51) --
	( 50.93, 51.93) --
	( 52.85, 52.34) --
	( 54.83, 52.73) --
	( 56.89, 53.12) --
	( 59.01, 53.49) --
	( 61.19, 53.86) --
	( 63.44, 54.21) --
	( 65.75, 54.54) --
	( 68.11, 54.87) --
	( 70.53, 55.18) --
	( 73.01, 55.47) --
	( 75.53, 55.75) --
	( 78.10, 56.02) --
	( 80.72, 56.27) --
	( 83.38, 56.51) --
	( 86.07, 56.73) --
	( 88.81, 56.94) --
	( 91.58, 57.13) --
	( 94.38, 57.30) --
	( 97.22, 57.46) --
	(100.07, 57.60) --
	(102.96, 57.73) --
	(105.86, 57.84) --
	(108.78, 57.93) --
	(111.71, 58.01) --
	(114.66, 58.07) --
	(117.61, 58.11) --
	(120.57, 58.14) --
	(123.54, 58.14);
\definecolor{fillColor}{gray}{0.70}

\path[fill=fillColor] ( 82.80, 25.85) circle ( 11);

\path[fill=fillColor] (164.27, 25.85) circle ( 11);

\path[fill=fillColor] (215.07, 37.37) circle ( 8.5);

\path[fill=fillColor] (196.94, 51.75) circle (  6);

\path[fill=fillColor] (123.54, 58.14) circle (  5);

\path[fill=fillColor] ( 50.13, 51.75) circle (  6);

\path[fill=fillColor] ( 32.00, 37.37) circle ( 8.5);
\definecolor{drawColor}{RGB}{0,0,255}

\path[draw=drawColor,line width= 1.2pt,line join=round,line cap=round] ( 82.80, 25.85) --
	( 83.57, 25.78) --
	( 84.34, 25.72) --
	( 85.12, 25.65) --
	( 85.90, 25.59) --
	( 86.68, 25.53) --
	( 87.46, 25.47) --
	( 88.25, 25.41) --
	( 89.04, 25.35) --
	( 89.83, 25.30) --
	( 90.63, 25.24) --
	( 91.43, 25.19) --
	( 92.23, 25.14) --
	( 93.03, 25.09) --
	( 93.84, 25.04) --
	( 94.65, 24.99) --
	( 95.46, 24.94) --
	( 96.27, 24.90) --
	( 97.09, 24.85) --
	( 97.91, 24.81) --
	( 98.73, 24.77) --
	( 99.55, 24.73) --
	(100.37, 24.69) --
	(101.20, 24.65) --
	(102.02, 24.62) --
	(102.85, 24.58) --
	(103.69, 24.55) --
	(104.52, 24.52) --
	(105.35, 24.49) --
	(106.19, 24.46) --
	(107.03, 24.43) --
	(107.86, 24.40) --
	(108.70, 24.38) --
	(109.55, 24.35) --
	(110.39, 24.33) --
	(111.23, 24.31) --
	(112.08, 24.29) --
	(112.92, 24.27) --
	(113.77, 24.26) --
	(114.61, 24.24) --
	(115.46, 24.23) --
	(116.31, 24.22) --
	(117.16, 24.20) --
	(118.01, 24.19) --
	(118.86, 24.19) --
	(119.71, 24.18) --
	(120.56, 24.17) --
	(121.41, 24.17) --
	(122.26, 24.17) --
	(123.11, 24.16) --
	(123.96, 24.16) --
	(124.82, 24.17) --
	(125.67, 24.17) --
	(126.52, 24.17) --
	(127.37, 24.18) --
	(128.22, 24.19) --
	(129.07, 24.19) --
	(129.92, 24.20) --
	(130.77, 24.22) --
	(131.62, 24.23) --
	(132.46, 24.24) --
	(133.31, 24.26) --
	(134.16, 24.27) --
	(135.00, 24.29) --
	(135.85, 24.31) --
	(136.69, 24.33) --
	(137.53, 24.35) --
	(138.37, 24.38) --
	(139.21, 24.40) --
	(140.05, 24.43) --
	(140.89, 24.46) --
	(141.72, 24.49) --
	(142.56, 24.52) --
	(143.39, 24.55) --
	(144.22, 24.58) --
	(145.05, 24.62) --
	(145.88, 24.65) --
	(146.71, 24.69) --
	(147.53, 24.73) --
	(148.35, 24.77) --
	(149.17, 24.81) --
	(149.99, 24.85) --
	(150.81, 24.90) --
	(151.62, 24.94) --
	(152.43, 24.99) --
	(153.24, 25.04) --
	(154.04, 25.09) --
	(154.85, 25.14) --
	(155.65, 25.19) --
	(156.45, 25.24) --
	(157.24, 25.30) --
	(158.04, 25.35) --
	(158.83, 25.41) --
	(159.62, 25.47) --
	(160.40, 25.53) --
	(161.18, 25.59) --
	(161.96, 25.65) --
	(162.73, 25.72) --
	(163.51, 25.78) --
	(164.27, 25.85);

\path[draw=drawColor,line width= 1.2pt,line join=round,line cap=round] ( 82.80, 25.85) -- ( 83.57, 25.78);

\path[draw=drawColor,line width= 1.2pt,line join=round,line cap=round] ( 83.24, 30.89) --
	( 82.80, 25.85) --
	( 82.37, 20.81);

\path[draw=drawColor,line width= 1.2pt,line join=round,line cap=round] (164.27, 25.85) -- (163.51, 25.78);

\path[draw=drawColor,line width= 1.2pt,line join=round,line cap=round] (164.71, 20.81) --
	(164.27, 25.85) --
	(163.84, 30.89);
\definecolor{drawColor}{RGB}{0,0,0}


\definecolor{drawColor}{RGB}{0,0,255}

\path[draw=drawColor,line width= 0.4pt,line join=round,line cap=round] ( 68.72, 41.15) -- ( 96.89, 41.15);

\path[draw=drawColor,line width= 0.4pt,line join=round,line cap=round] ( 93.76, 39.35) --
	( 96.89, 41.15) --
	( 93.76, 42.96);
\definecolor{drawColor}{RGB}{0,0,0}

\node[text=drawColor,anchor=base,inner sep=0pt, outer sep=0pt, scale=  1.1] at ( 82.80, 44.25) {$v_n$};

\definecolor{drawColor}{RGB}{0,0,255}

\path[draw=drawColor,line width= 0.4pt,line join=round,line cap=round] (150.19, 41.15) -- (178.36, 41.15);

\path[draw=drawColor,line width= 0.4pt,line join=round,line cap=round] (175.23, 39.35) --
	(178.36, 41.15) --
	(175.23, 42.96);
\definecolor{drawColor}{RGB}{0,0,0}

\node[text=drawColor,anchor=base,inner sep=0pt, outer sep=0pt, scale=  1.1] at (164.27, 44.25) {$v_{n+1}$};

\node[text=drawColor,anchor=base,inner sep=0pt, outer sep=0pt, scale=  1.1] at (123.54, 29) {$x_{n+1}-x_n$};

\node[text=drawColor,anchor=base,inner sep=0pt, outer sep=0pt, scale=  1] at ( 35, 17) {Ring of length} ;

\node[text=drawColor,anchor=base,inner sep=0pt, outer sep=0pt, scale=  1] at ( 35, 10) {$L=231$~m} ;

\node[text=drawColor,anchor=base,inner sep=0pt, outer sep=0pt, scale=  1] at (215, 13.45) {$N=22$ vehicles};

\node[text=drawColor,anchor=base,inner sep=0pt, outer sep=0pt, scale=  1] at ( 82.80, 7) {$n$-th vehicle};

\node[text=drawColor,anchor=base,inner sep=0pt, outer sep=0pt, scale=  1] at (164.27, 7) {$(n+1)$-th vehicle};

\end{scope}
\end{tikzpicture}
    \caption{Replica of the experiment by Sugiyama et al.\ \cite{sugiyama2008traffic}.}
    \label{fig:scheme}
\end{figure}
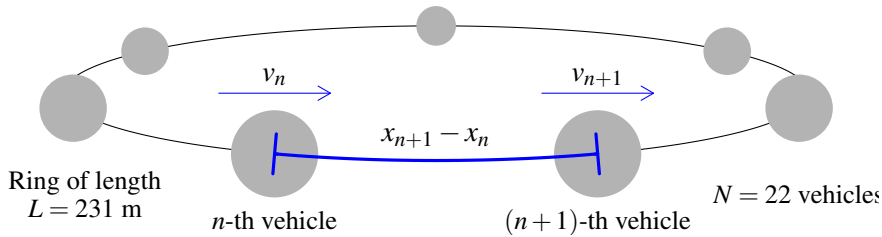

In the following, we simulate $N=22$ vehicles on a single-lane circuit of length $L=231$~m with periodic boundaries to mimic the conditions of the experiment by Sugiyama et al.~\cite{sugiyama2008traffic} (see Fig.~\ref{fig:scheme}). 
The simulations are performed using the Euler explicit/implicit numerical scheme \eqref{eq:SATGdiscrete} of the  stochastic ATG model \eqref{eq:SATG}.
An online simulation module is available at: 
    \url{https://www.vzu.uni-wuppertal.de/fileadmin/site/vzu/Noise-Induced_Stop-and-Go_Modelling_and_Control.html?speed=0.7}.

\begin{figure}[!ht]
    \centering\footnotesize\medskip
    \includegraphics[width=.8\textwidth]{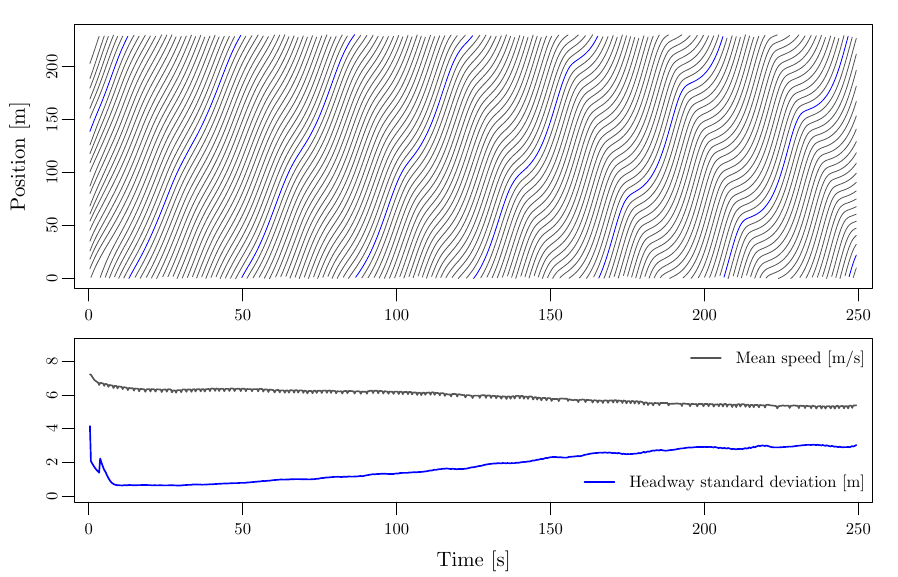}
    \caption{Simulated trajectories of 22 vehicles on a 231-metre circuit as in the experiment by Sugiyama et al. \cite{sugiyama2008traffic} using the stochastic ATG model \eqref{eq:SATGdiscrete} where $\sigma=2.8$~m. The trajectories can be simulated online at: 
    \protect\url{https://www.vzu.uni-wuppertal.de/fileadmin/site/vzu/Noise-Induced_Stop-and-Go_Modelling_and_Control.html?speed=0.7}.}
    \label{fig:TrajSim}  
\end{figure}

\paragraph{Trajectories from a single simulation}
Fig.~\ref{fig:TrajSim} shows an example of simulated trajectories over 250 seconds. 
The initial vehicle positions and speeds are the same as in the experiment by Sugiyama et al. \cite{sugiyama2008traffic}. 
The noise amplitude is $\sigma=2.8$~m. 
A single stop-and-go wave emerges after approximately two minutes, as observed in the experiment, compare Figs.~\ref{fig:TrajExp} and \ref{fig:TrajSim}. 
Additionally, the sequences of mean speed and gap standard deviation exhibit similar trends to those observed in the experiment.
However, the trajectories and mean sequences appear much more regular, primarily because all vehicles behave identically in the model.
In fact, there is no heterogeneity in driving style that we would expect from real drivers. 
Furthermore, the noise affects only the distance variable, which is an input smoothed by the model. 
Adding noise directly to the acceleration makes the trajectories less regular, see \cite[Fig.~4]{dufour2025noiseinducedtransitionstopandgowaves}.

\begin{figure}[!ht]
    \centering\bigskip
    \input{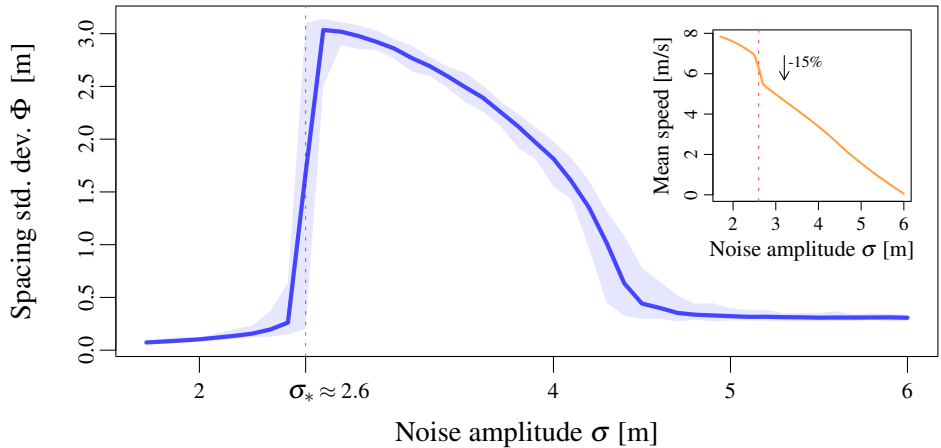}
    \caption{Gap standard deviation for the 22 vehicles on the 231~m circuit (replica of the Sugiyama experiment) with the noisy ATG model \eqref{eq:SATG} obtained from variation of the noise amplitude. The continuous lines are the averaged gap standard deviations of $K=100$ simulations, while the coloured areas are the min/max ranges. A phase transition arises from stable uniform solutions to stop-and-go dynamics as the noise amplitude increases.}
    \label{fig:PhaseTransition}  
\end{figure}

\paragraph{Noise-induced transition to stop-and-go dynamics }
Several independent simulations are performed by varying the noise amplitude $\sigma$ from 1 to 6~m in steps of $0.1$~m. All the other parameters $\tau=5$~s, $ T=0.9$~s, $\ell=3$~m, $\tmin=0.1$~s, and $\tmax=2.5$~s remain constant. 
We repeat the simulations $K=100$ times for each value of the noise amplitude, starting from uniform initial conditions. 
The simulations are run for  $t_S=1\,000$~seconds to reach a stationary state, before we average over time the spacing standard deviation
\begin{equation}
    \Phi(t_S,t_M)=\frac{\delta t}{t_M}\sum_{t=t_S}^{t_S+t_M}\sqrt{\frac{1}{N-1}\sum_{n=1}^N \bigl( \Delta x_n(t)-L/N\bigr)^2}
\end{equation}
during the next $t_M=100$~seconds. 
The results are shown in Fig.~\ref{fig:PhaseTransition}. 
A phase transition is clearly observed when the noise amplitude exceeds approximately $\sigma_\ast\approx2.6$~m. 
Additionally, the mean speed drops by around 15\% as the stop-and-go dynamics emerge (see the subplot in Fig.~\ref{fig:PhaseTransition}). 
A similar capacity reduction is also evident in the experiment, see Fig.~\ref{fig:TrajExp} and \cite[Fig.~6.6]{Schadschneider:1564539}. 
This decrease of the mean speed is due to the nonlinear form of the ATG model and its asymmetric behaviour during braking and acceleration stop-and-go phases. 
Interestingly, an optimal noise amplitude for the emergence of waves can be identified as around $\sigma\approx 2.7$~m. 
Beyond this optimal noise amplitude, the spacing standard deviation decreases progressively until it reaches a low constant for values of $\sigma$ beyond  4.5~meters, meaning that no stop-and-go waves arise. 

\paragraph{Sensitivity analysis -- Influence of the model's parameters on stability}

\begin{figure}[!ht]
    \centering\smallskip
    \input{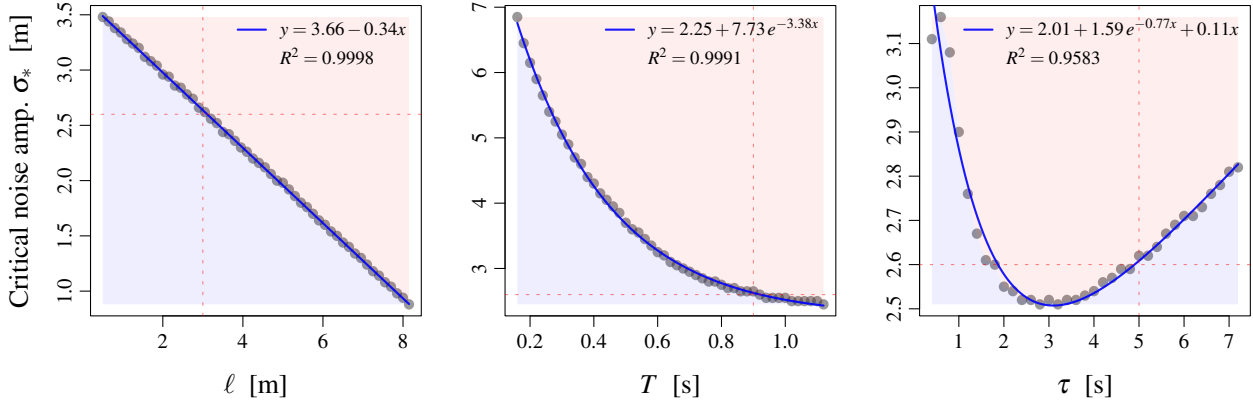}
    \caption{Critical noise amplitude threshold according to the three main parameters of the ATG model (from left to right): vehicle size $\ell$, time gap $T$, and sensitivity $\tau$. The system is unstable and waves propagate above the critical curve (red part) while the system is stable below it (blue part).}
    \label{fig:ParaSpace}
\end{figure}

The critical noise amplitude threshold for wave formation depends on the density level and the values of the model parameters.
Increasing the values of the parameters $\ell$ and $T$, i.e.\ reducing the mean vehicle speed, results in stop-and-go dynamics emerging at a lower noise amplitude.
More precisely, a linear relationship can be identified between the parameter $\ell$ controlling the mean gap and the critical noise amplitude, see Fig.~\ref{fig:ParaSpace}, left panel. 
This result is expected since the noise is additive and affects the gap, i.e. a distance variable.
In fact, the waves emerge when the noise amplitude is higher than around 35\% of the mean gap.
A nonlinear exponentially decreasing relationship occurs with the time gap parameter $T$, see Fig.~\ref{fig:ParaSpace}, middle panel. Surprisingly, reducing the time gap, i.e., increasing the mean vehicle speed, improves stability. 
The relationship with the relaxation time parameter $\tau$ is more complex. 
It is nonmonotonic, see Fig.~\ref{fig:ParaSpace}, right panel. 
The stability of the model is most critical at approximately $\tau=3$. 
Beyond this point, more reactive or smoother behaviour improves stability.

\paragraph{Noise driven by the Ornstein-Uhlenbeck process}

\begin{figure}[!ht]
    \centering\bigskip
    \input{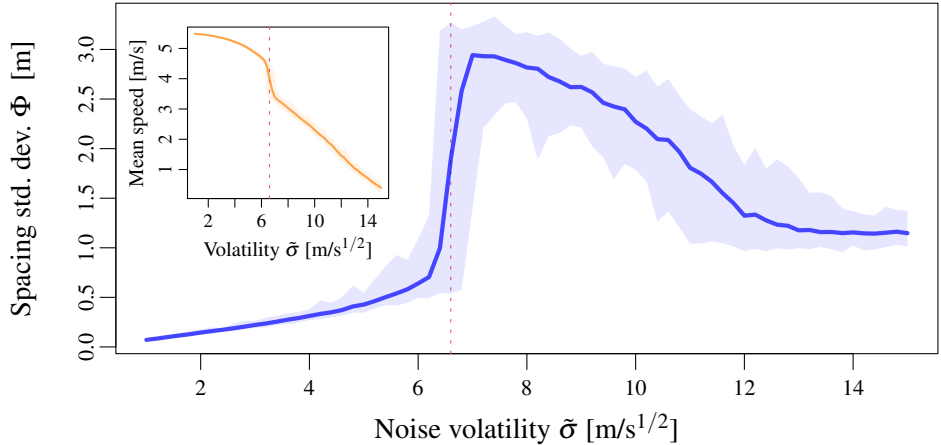}
    \caption{Gap standard deviation for the 22 vehicles on the 231~m circuit (replica of the Sugiyama experiment) with the noisy ATG model \eqref{eq:SATG_OU} driven by the Ornstein-Uhlenbeck process. The noise volatility $\tilde\sigma$ vary from 0 to 15~m/s$^{1/2}$. Here, $\beta=5~s$ to introduce a large time-correlation of the noise. The continuous lines are the averaged gap standard deviations of $K=100$ simulations, while the coloured areas are the min/max ranges. As with the white noise model, a phase transition arises from stable uniform solutions to stop-and-go dynamics as the noise amplitude increases, compare with Figure~\ref{fig:PhaseTransition}.}
    \label{fig:PhaseTransition_OU}
\end{figure}

For modelling purposes and sake of simplicity, it is convenient to assume that white noise affects the measurement of the gap.
However, human and vehicle perception is not atomistic. 
It would be more realistic to assume that measurement perturbations are correlated over time. In this section, we carry out further simulations using independent Ornstein-Uhlenbeck processes, denoted by $\tilde\xi_n$, instead of the white noise $\sigma\xi_n(t)$ in \eqref{eq:SATG}. 
The ATG model driven by the Ornstein-Uhlenbeck process is given by
\begin{equation}
\left\{~\begin{aligned}
    \mathrm{d}v_n(t) &=\frac{\frac1\tau \Big[g_n(t)+\tilde\xi_n(t)- Tv_n(t)\Big]+\Delta v_n(t)}{T_\varepsilon\big(g_n(t)+\tilde\xi_n(t),v_n(t)\big)}\mathrm{d}t,\\
    \mathrm{d}\tilde\xi_n(t)&=-\frac1\beta\tilde\xi_n(t)\text d t+\tilde\sigma dW_n(t),
    \end{aligned}\right.
    \label{eq:SATG_OU}
\end{equation}
where $W_n$ are standard Wiener processes.
The Ornstein-Uhlenbeck process $\tilde\xi_n$ is a Gaussian process with asymptotic expected value zero and variance $0.5\beta\tilde\sigma^2$.
Unlike the white noise model, which is uncorrelated over time, the time correlation of the Ornstein-Uhlenbeck process is 
$ \langle\tilde\xi_n(t),\tilde\xi_n(t')\rangle = \frac12\beta\tilde\sigma^2 e^{-|t-t'|/\beta}.$
We set the relaxation time $\beta=5$~s to introduce a large time correlation of the noise.
A similar phase transition can be observed from about $\tilde\sigma>6.5$~m/s$^{1/2}$, see Fig.~\ref{fig:PhaseTransition_OU} and compare with Fig.~\ref{fig:PhaseTransition}. 
Further simulation results show that comparable noise-induced oscillations occur when the noise is added to the acceleration, to the speed difference, or to a combination of the acceleration, the speed difference, and the gap.
In \cite{dufour2025noiseinducedtransitionstopandgowaves}, white and state-dependent noises that vanish when the speed approaches zero are introduced into the acceleration function of the ATG model, yielding similar qualitative phase transitions.
Therefore, nature of the noise does not appear to be the primary factor in the phase transition.
The most important aspect is that the system is perturbed stochastically.

\section{Stabilisation of the dynamics\label{Sec:Stab}}
In real-world scenarios, uncertainties regarding the gap and velocity of the leading vehicle present significant challenges for ACC systems. The ACC controller must ensure that these uncertainties do not result in excessive acceleration and braking events, which could lead to an uncomfortable or even unsafe driving experience~\cite{maruyama2022proposal}. To address this, ACC systems are designed to be less reactive, incorporating filtering and smoothing mechanisms. The treatment of the resulting dilemma between comfortable driving experience and traffic stability is still an open research question~\cite{winner2009adaptive}.

\begin{figure}[!ht]
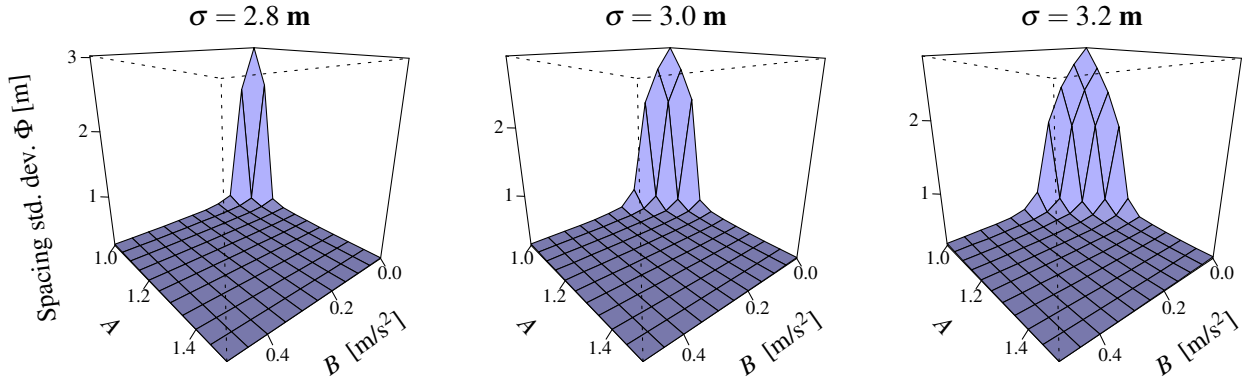

\bigskip
    \hspace{5mm}\input{Figs/SD_sig2.8}\input{Figs/SD_sig3}\input{Figs/SD_sig3.2}\\[-45mm]
    \rotatebox{97}{\footnotesize Spacing std.~dev.~$\Phi$ [m]}\\[7mm]
    \caption{Average spacing standard deviation for 22 vehicles on a 231~m circuit (replicating the Sugiyama experiment) using the stochastic ATG model \eqref{eq:SATGab}, with varying scale factor $A$ and acceleration bias $B$. The spacing deviation approaches zero, indicating wave dissipation, as $A$ and $B$ increase.}\medskip
    \label{fig:Dissipation}  
\end{figure}

To address the issue of instability within the stochastic Adaptive Time Gap  model, we propose in this section a minimalist affine transformation of the system dynamics. This approach aims to suppress noise-induced nonlinear instabilities of the uniform solution. The affine-transformed model is defined by
\begin{equation}
    \mathrm{d}v_n(t) = \left[A\,\frac{\frac1\tau \left(g_n(t)+\sigma\xi_n(t)-T v_n(t)\right)+\Delta v_n(t)}{T_\varepsilon\big(g_n(t)+\sigma\xi_n(t),v_n(t)\big)}+B\right]\mathrm{d}t,\qquad A>0,~B\in\mathbb R,
    \label{eq:SATGab}
\end{equation}
where $A$ represents a scale factor and $B$ an acceleration bias. Accordingly, the partial derivatives of the linearized deterministic model at equilibrium are obtained by scaling the expressions in \eqref{ATG_linearized_PD} by the factor $A$. From \eqref{eq:stable}, the linearized deterministic system fulfills the general stability condition when 
\begin{equation} 
    A \ge \frac{2\tau}{T+2\tau}\,. \label{A_string_stability} 
\end{equation}
Furthermore, using \eqref{eq:ovdstable}, the system satisfies the overdamped stability criterion simply when $A \ge 1$, which also guarantees~\eqref{A_string_stability}.
Specifically, \( A > 1 \) amplifies the control response to quickly counteract noise effects while \( B > 0 \) introduces a bias towards acceleration to prevent excessive deceleration and avoid wave amplification. The simulations demonstrate that for sufficiently large values of \( A \) and \( B \) the system can effectively dissipate waves even under high noise amplitudes. As shown in Fig.~\ref{fig:Dissipation}, when $A \geq 1.2 $ and $ B \geq 0.2  m/s^2$, the spacing standard deviation in stationary states approaches zero even for high values of the noise amplitudes $ \sigma $, indicating successful wave dissipation.


\begin{figure}[!ht]
    \centering\bigskip
    \begin{subfigure}{0.49\textwidth}
        \centering
        \includegraphics[width=\textwidth]{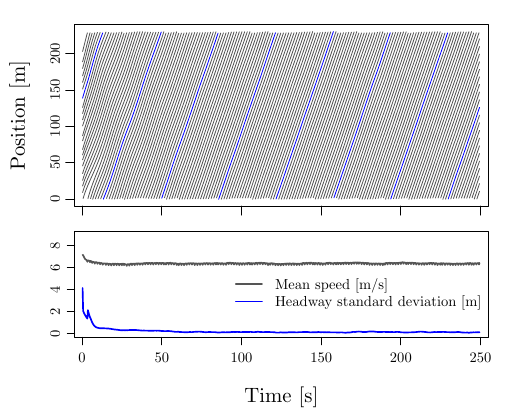}\medskip
    \end{subfigure}
    \hfill
    \begin{subfigure}{0.49\textwidth}
        \centering
        \includegraphics[width=\textwidth]{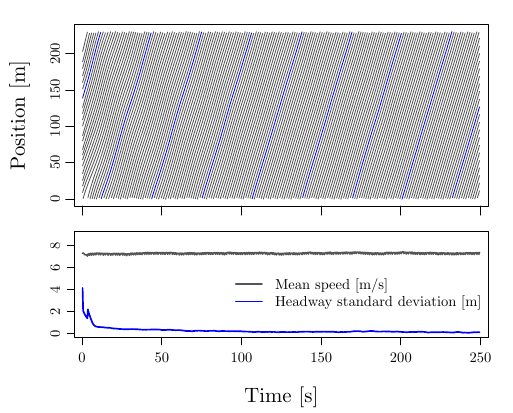}\medskip
    \end{subfigure}
    \caption{Simulated trajectories of 22 vehicles on a 231-metre circuit using the extended stochastic ATG model \eqref{eq:SATGab} with \( \sigma = 2.8 \) m and where \( A = 1.2 \), \( B = 0 \) m/s\(^2\) (left panels), and \( A = 1 \), \( B = 0.2 \) m/s\(^2\) (right panels). Compare to the untable baseline model in Fig.~\ref{fig:TrajSim}.}
    \label{fig:SimFineTunedCombined}
\end{figure}

The beneficial effects of this transformation are further illustrated in Fig.~\ref{fig:SimFineTunedCombined} where we replicate the experiment from Fig.~\ref{fig:TrajSim}. Keeping all conditions identical, we introduce the tuning parameters $A = 1.2$, \( B = 0 \)~m/s\(^2\) in Fig.~\ref{fig:SimFineTunedCombined}~(a) and the combination \( A = 1 \), \( B = 0.2 \) m/s\(^2\) in Fig.~\ref{fig:SimFineTunedCombined}~(b). In both cases, stop-and-go waves are entirely eliminated. 
In addition, the speed is slightly increased when the bias $B$ is strictly positive, see Fig.~\ref{fig:SimFineTunedCombined}~(b), grey curve in the lower panel. 
Specifically, a positive acceleration bias $B$ increases the system's equilibrium speed according to
\begin{equation}
     v^*(g)=\frac{g}T+\frac{B\tau}{A},
\end{equation}
for any given gap $g\ge0$. 
As a result, larger values of $B$, longer relaxation times $\tau$, or smaller factor $A$ all contribute to an increase in the equilibrium speed relative to the initial equilibrium value $g/T$.

While the affine modification \eqref{eq:SATGab} improves stability, it is not without trade-offs. High values of the factor $A$ can result in uncomfortable levels of acceleration and a positive bias $B$ may compromise safety by increasing the speed and so reducing the time gap to the lead vehicle. Specifically, the resulting equilibrium time gap becomes:
\begin{equation}
     T^*=T\Big(1 - \frac{B\tau} {Av^*}\Big)=\frac{gT}{g+T\frac{B\tau} {A}}.
\end{equation}
When $B\ge0$, this effective time gap \( T^* \) decreases as speed \( v^* \) or gap \( g \) approaches zero and asymptotically tends to the nominal time gap \( T \) at higher speeds or gaps. However, it remains always smaller than the nominal gap. From a safety standpoint, this can be problematic at low speeds due to reduced headway and should therefore be bounded by a minimal nominal time gap. However, from a performance perspective, especially in vehicle platooning, this behaviour is advantageous. Reduced time gaps typically enhance throughput and efficiency. Interestingly, in contrast to traditional approaches, where stability improves at the cost of increased time gaps, this method offers an alternative solution. By allowing \( T^*\) to increase with speed, the system maintains safety at higher velocities by mitigating noise-induced instabilities.

\section{Discussion\label{Sec:Dis}}


The simulation results demonstrate that oscillatory stop-and-go waves caused by noise occur with many different types of noise in the ATG car-following model. 
These include white noise and time-correlated noise, as well as noise acting on the gap, speed difference or acceleration \cite{dufour2025noiseinducedtransitionstopandgowaves}.
The uniform solution is systematically stable for small perturbations but becomes unstable for large ones, resulting in stop-and-go dynamics. 
The ATG model is unconditionally linearly stable. 
In fact, the uniform solution is stable for all initial conditions, whether they are jam wave or simply random configurations. 
Therefore, stop-and-go dynamics result from nonlinear instability purely induced by noise. 
This phenomenon is analogous to Kapitza's pendulum \cite{Kapitza1951}, where the stable solution switches to an inverted pendulum when perturbed.
The noise causes the system to oscillate at the largest wavelength.
Although this hidden frequency is unstable, it represents the most stable oscillatory configuration of the system.
Indeed for the ATG model, the root of the characteristic equation \eqref{eq:EquaCharac} with the largest real part is that for $\theta\to0$. 

The noise causes the system to switch from a laminar state to an oscillatory state with stop-and-go dynamics.
In addition, an optimal noise amplitude can be identified for the emergence of waves.
These are characteristic features of stochastic resonance in bistable systems \cite{gammaitoni1998stochastic}.
However, in this study, the noise directly affects the stationary states of the system and their stability. 
The uniform equilibrium is stable in the deterministic case. 
The system converges to laminar states regardless of the initial conditions.
Under large perturbations, the uniform equilibrium is unstable and the oscillatory solution with stop-and-go waves becomes stable instead.
Nevertheless, if the perturbed system reaches an oscillatory configuration with stop-and-go waves, the flow will become laminar again if the noise disappears.
This behaviour differs from that of classical stochastic resonance, where noise merely facilitates transitions between two locally stable equilibrium states \cite{benzi1981mechanism,fauve1983stochastic}. 
Here, the transition to oscillatory dynamics is purely induced by noise and could be interpreted as a form of stochastic stabilization \cite{appleby2005stochastic} since the noise enables oscillatory solutions with stop-and-go waves to become stable.
However, unlike in classical stochastic stabilisation phenomena, the stabilised solution is not a stationary equilibrium but rather an oscillatory collective pattern in response to perturbations.

The proposed linear transformation of the stochastic dynamics in \eqref{eq:SATGab} provides a simple yet effective means of counteracting noise-induced destabilisation. Stability can be recovered either by amplifying the model response (slope parameter $A$) or by introducing a positive acceleration bias (intercept parameter $B$). From a physical perspective, this transformation plays a role analogous to classical linear stabilisation mechanisms in dynamical systems. Amplification of the response modifies the effective restoring forces, similarly to how parametric amplification or feedback gains can suppress fluctuations in noisy oscillators. Likewise, the addition of a constant bias term is reminiscent of tilting an effective potential landscape, thereby favouring stable configurations and suppressing transitions induced by stochastic forcing. In this sense, the transformation acts as a control-induced reshaping of the system’s stability properties rather than altering its nonlinear structure. This parallels stabilisation strategies encountered in classical mechanics and statistical physics, where linear feedback or biasing fields are used to stabilise noise-sensitive states. Applied to traffic flow, this result highlights that minimal, linear modifications of vehicle response, either through increased sensitivity or systematic acceleration bias, can robustly mitigate noise-induced stop-and-go waves and restore uniform flow.

\section{Conclusion\label{Sec:Ccl}}

Gaussian noise in the nonlinear ATG car-following model can lead to the emergence of oscillatory dynamics and stop-and-go waves, although the model is unconditionally linearly stable and the noise is white. 
In fact, the uniform solution remains stable for small perturbations, but becomes unstable and gives way to stop-and-go dynamics for large perturbations. 
This applies to many different types of noise, including white noise and time-correlated noise, as well as noise acting on the gap, speed difference, or acceleration.
Similar to Kapitza's pendulum in a stochastic framework, the noise in the nonlinear dynamics affects the stability of the system and induces a phase transition including a metastable regime. 
In this stochastic approach, the waves result from a nonlinear instability in contrast to classical delayed and inertial car-following models where the transition is based on a linear instability. 
Intriguingly, a simple affine transformation of the model allows the waves to dissipate  and the stability of uniform solutions to be recovered. 
Similar results have been observed using heterogeneous car-following models \cite{ehrhardt2024stability}. 
Amplifying the response and adding a positive bias seems to be an effective strategy to mitigate the instabilities induced by the noise and other heterogeneity factors. 

Beyond improving stability, the affine transformation of the model has notable implications. 
Large factors $A$ can cause excessive acceleration, while positive biases $B$ may compromise safety by reducing the effective time gap, especially at low speeds. 
This equilibrium time gap increases with speed but remains below the desired time gap $T$, thus potentially posing safety concerns. 
However, this behaviour enhances performance—particularly in platooning—since stability improves with reduced time gaps. 
Unlike traditional approaches where increasing the time gap boosts stability at the cost of efficiency, this method leverages a speed-dependent time gap to mitigate noise-induced nonlinear instability while maintaining high performance.
It would be interesting to investigate whether the affine transformation of the model could also compensate for the linear instabilities of delayed or inertial models.
Preliminary simulation results show that this is not the case for the delayed or lag ATG model. 
This is not surprising. 
Although the stop-and-go waves in the case of a linear instability due to delay and inertia or a nonlinear instability due to noise are qualitatively similar, the mechanisms involved are different. 
Specific compensators are therefore required to reduce delays in the dynamics.
Unified algorithms need to be developed that are robust to both stochastic perturbations and delays in the dynamics of the system.

\section*{Acknowledgement}
RK and AT acknowledge the German Research Foundation (Deutsche Forschungsgemeinschaft - DFG) for funding this research, grant number 546728715.

\bibliographystyle{abbrv}
\setlength{\bibsep}{7pt}
\bibliography{reference}

\end{document}